\documentclass{llncs}
\usepackage{epsfig}
\usepackage{calc}
\usepackage{multicol}
\usepackage{pslatex}
\usepackage{apalike}
\usepackage{epsfig}
\usepackage{rotating}
\usepackage{url}
\usepackage{pifont}% http://ctan.org/pkg/pifont
\newcommand{\cmark}{\ding{51}}%
\newcommand{\xmark}{\ding{55}}%
\usepackage{comment}
\begin{document}

%openin

\title{Deconstructing the Decentralization Trilemma}
\titlerunning{Decentralization Trilemma}

\authorrunning{Halpin} % abbreviated author list (for running head)

\author{Harry Halpin}
  \institute{  
INRIA, 2 Simone Iff 75012,\\
Paris, France \\
\email{harry.halpin@inria.fr} \\
}

%\begin{document}

%\author{\authorname{Harry Halpin\sup{1}\orcidAuthor{0000-0003-2143-6965}}
%\affiliation{\sup{1}Inria de Paris, 2 Rue Simone Iff, Paris, France}
%\email{harry.halpin@inria.fr}
%}

%\keywords{software architecture, security, privacy, scalability, federation, decentralization, blockchain}

%\onecolumn \maketitle \normalsize \setcounter{footnote}{0} \vfill

  \maketitle

\begin{abstract} The vast majority of applications at this moment rely on centralized servers to relay messages between clients, where these servers are considered trusted third-parties.  With the rise of blockchain technologies over the last few years, there has been a move away from both centralized servers and traditional federated models to more decentralized peer-to-peer alternatives. However, there appears to be a trilemma between security, scalability, and decentralization in blockchain-based systems. Deconstructing this trilemma using well-known threat models, we define a typology of centralized, federated, and decentralized architectures. Each of the different architectures has this trilemma play out differently. Facing a possible decentralized future, we  outline seven hard problems facing decentralization and theorize that the differences between centralized, federated, and decentralized architectures depend on differing social interpretations of trust.
\end{abstract}
  
\section{Introduction}

Although there has been a move towards decentralization, projects with decentralized architectures have had fundamental difficulties: Bitcoin and Ethereum seem to be unable to scale to as large a number of transactions as centralized systems such as Visa. On the other hand, centralized projects are increasingly the subject of data leakage and other attacks, calling their security into question. This has been phrased as the ``decentralization trilemma'' by the co-founder of Ethereum, Vitalik Buterin, and is a widely spread truism in blockchain development that has not been rigorously analyzed and critiqued.\footnote{\url{https://github.com/ethereum/wiki/wiki/Sharding-FAQ}}

\begin{figure}
\centering
 \includegraphics[scale=0.4]{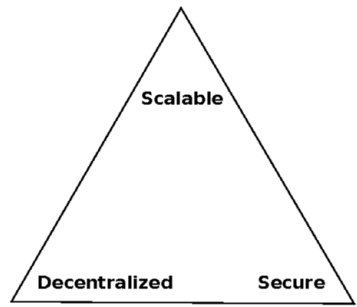}
\caption{The decentralization trilemma}
\label{fig:triangle}
\end{figure}

Intuitively, there does seem to be fundamental trade-offs between decentralization, scalability, and security as illustrated in Figure \ref{fig:triangle}: Systems can be less decentralized and more scalable versus more decentralized and less scalable.  Indeed, decentralization does seem like a trade-off against scalability, but most large real-world deployments of scalable software, such as Amazon, are actually distributed systems with large trust assumptions and centralized co-ordination~\cite{dynamo}. Also, there are secure centralized systems that use advanced encryption (such as the Signal instant messenger), and decentralized systems that have been found to be insecure, as various attacks on the distributed hash tables used by peer-to-peer file-sharing networks show~\cite{WolchokHHFHRWW10}. Decentralization can be thought of as arising from a separate threat model than traditional security assumptions: A lack of trust in centralized servers.  In this paper, we outline the threat model, the \emph{malicious server}, that these decentralized architectures are trying to address in Section ~\ref{sec:threat}. Decentralization, which we define in terms of an adversarial approach to distributed systems~\cite{troncoso2017systematizing}, is then explored in Section ~\ref{sec:properties} as separate from classical security and scalability requirements. Rather than a binary division, we view decentralization on a spectrum that can be broadly construed as  centralized, federated, and decentralized architectures in Section \ref{sec:arch}, and we provide an analysis in Section \ref{sec:analysis}. In Section \ref{sec:problems} we outline six open problems that decentralized systems face in meeting the requirements currently met by centralized architectures. In our conclusion in Section~\ref{sec:end}, we reassess decentralization and turn to the social hypothesis at the heart of decentralization.

\section{The Malicious Server}
\label{sec:threat}

In distributed systems, all entities are considered to be capable of sending messages~\cite{lamport1982byzantine}.  In centralized systems, users do not directly receive messages but are mediated by a client (a device, a program such as a mail-reader or a browser, etc.) where  the client communicates to a server that stores and forwards messages to the client. As exemplified by cloud computing, this server is assumed to be always online. One of the primary advantages of centralized servers seems to be that the deployment and upgrading of any protocol is easier via the usage of a centralized server.

As the server mediates all messages, the server is a \emph{trusted third party}. Centralized systems can be secure and maintain privacy against even powerful adversaries. Secure messaging applications ranging from Signal to WhatsApp, depend on centralized servers~\cite{unger2015sok}. In the case of secure messaging protocols, even if the message content is encrypted, the server is usually necessary for delivery, especially if the client is offline. As shown by the simple case that messages are assumed not be dropped by the server, the security and privacy properties of centralized servers are dependent on a single root of trust.

Under the adversary model of \emph{malicious security}, trusted components of a larger distributed process may no longer follow the protocol and so no longer maintain their security properties~\cite{yung2015mobile}. To achieve malicious security, the protocol must work even with a component no longer maintains its security properties. Although originally aimed at cryptographic primitives,  malicious security also applies at the level of the entire architecture of a system. In particular, a malicious server is one that no longer securely relays messages according to the protocol.

We theorize that the goal of decentralization is to achieve protocols with malicious security in the face of any component, and so also achieve resistance in terms of the \emph{byzantine failure} of components, where  components behavior is unrestricted, and so may be malicious or ``unintentionally'' faulty (such as being offline)~\cite{lamport1982byzantine}. Given the popular deployment of servers in computing protocols, amongst all possible components, the primary threat model of decentralization is a \emph{malicious server}, where the server is untrusted. Malicious security can be achieved in a decentralized protocol simply by not relying on a server at all.  Malicious servers are a real-world adversary: As the revelations by Edward Snowden showed that it was trivial to compromise the security of e-mail if they were not end-to-end encrypted by simply retrieving the cleartext from the centralized server. Even if the message content is encrypted, ``metadata'' such as the time and recipients of the message, often has no legal protection from surveillance~\cite{slobogin2014cause}.

The goal of a malicious server in the centralized client-server setting is to read messages and possibly forge the message content of one or more clients. Even if the message is encrypted, the malicious server may be ``honest but curious'' and so also has the goal of determining the identity of the senders and recipients of messages, i.e. the discovery of the ``social graph'' of the communication. The server may use techniques like dropping messages, replaying messages, or sending fake messages to identify users. The server is \emph{local}, it does not have the ability to observe the entire network, but can observe every message it relays, and so differs from  threat models like the global passive adversary that can observe all messages in an entire network or an adversary that can arbitrarily corrupt any component in the system. The malicious server may be a mobile adversary and so the corruption of the server may be limited in duration~\cite{yung2015mobile}.

Before cloud computing, servers were generally viewed as trustworthy and message-passing between them governed primarily by standardized protocols such as SMTP and HTTP, leading to federated systems. Although the servers were generally regarded as trusted rather than malicious in early federated internet architectures, issues such as spam and a need for payments led to increased centralization. The pendulum is now swinging in the other direction: It is our fundamental hypothesis that decentralized systems arose in response to the threat model of the malicious server. For example, Bittorrent came into usage after the the centralized index of Napster was eliminated, and Bitcoin became popular after the centralized servers in classical e-cash schemes were viewed as a liability after the ``take down'' of e-Gold~\cite{troncoso2017systematizing}.

\section{Properties: Security and Scalability }
\label{sec:properties}

Security and scalability are very broad notions. It  is useful to decompose them into more distinct and better known technical properties that can be enforced via cryptography and engineering. The list of properties below is somewhat arbitrary, but reflects classical distinctions within the fields of computer security and distributed systems. 

\subsection{Security Properties}

A number of traditional of security properties are needed to counter a malicious server. Security properties should be able to hold in terms of messaging even in the presence of a malicious server, which is considered the ``adversary'' below: 

\begin{itemize}
\item \emph{Confidentiality}: The adversary cannot read the cleartext of any message.
\item \emph{Integrity}: The adversary cannot alter the message without detection.
\item \emph{Authenticity}: Only the intended recipient can receive and send the message (in order to prevent replay and impersonation attacks by the server).
\end{itemize}

Although informally thought of as ``security'' properties in the decentralization trilemma, we can consider privacy properties separately: 

\begin{itemize}
\item \emph{Unlinkability}: Each message of a client is unlinkable from any given other message by the client or any other client~\cite{PfitzmannH05}.
\item \emph{Unobservability}: A message cannot be distinguished from not sending a message~\cite{PfitzmannH05}.
\end{itemize}

\subsection{Scalability Properties}

Additional properties relate to the scalability of the system, and so can be contrasted with security and privacy requirements~\cite{das2018anonymity}. A system should be able to scale so it is both available to clients when needed, can support new clients, and deliver messages within a period of time acceptable to the user.
\begin{itemize}
\item \emph{Availability}: Messages can be sent or received by the client when requested.
\item \emph{Capacity}: New clients may be added at any time while maintaining availability.
\item \emph{Latency}: Messages may be sent by one client and received by another client within an acceptable period of time. 
\end{itemize}

Some requirements cannot be fulfilled by traditional cryptographic and engineering requirements, as these related to the freedom of users, particularly when they are under attack by a malicious server.

\begin{itemize}  
\item \emph{Transparency}: Messages or information about message-passing behavior are recorded for clients to access.
\item \emph{Portability}: Messages may be send and received across all servers and clients, so a user may move their messages to another server or send the message directly themselves in case their server becomes malicious.
\end{itemize}

These properties are by no means exhaustive, but are meant to deconstruct and ground the terms ``security'' and ``scalability'' in the decentralization trilemma in well-accepted technical properties. There are more informal properties that can fit within this framework. Take \emph{usability}, which can be defined as the ability to easily retrieve and send messages by the user.

\section{Decentralized Architectures}
\label{sec:arch}

In order to minimize the damage a malicious server could cause, decentralization is put forward as an alternative architecture to trusted third parties. However, there are multiple variations of architectures claiming to be decentralized: For example, efforts like Autocrypt claim that e-mail, despite involving a SMTP server, is decentralized.\footnote{\url{https://autocrypt.org/}} So defining and comparing these architectures in terms of messaging passing between clients and servers is necessary. We divide the three primary architectural choices as centralized, federated, and decentralized.  These architecture were classically illustrated by Baran in the following Figure \ref{fig:baran}~\cite{baran1964distributed}.\footnote{In his original work, Baran called these (1) Centralized, (2) Decentralized, and (3) Distributed, but terminology has been changed to be consistent with current software architecture practice where ``federated'' has a clear meaning in terms of a multiple authoritative servers with one or more clients each, while the difference between ``decentralized'' and ``distributed'' in informal modern parlance has become vague.}

\begin{figure}
\centering
 \includegraphics[width=\linewidth]{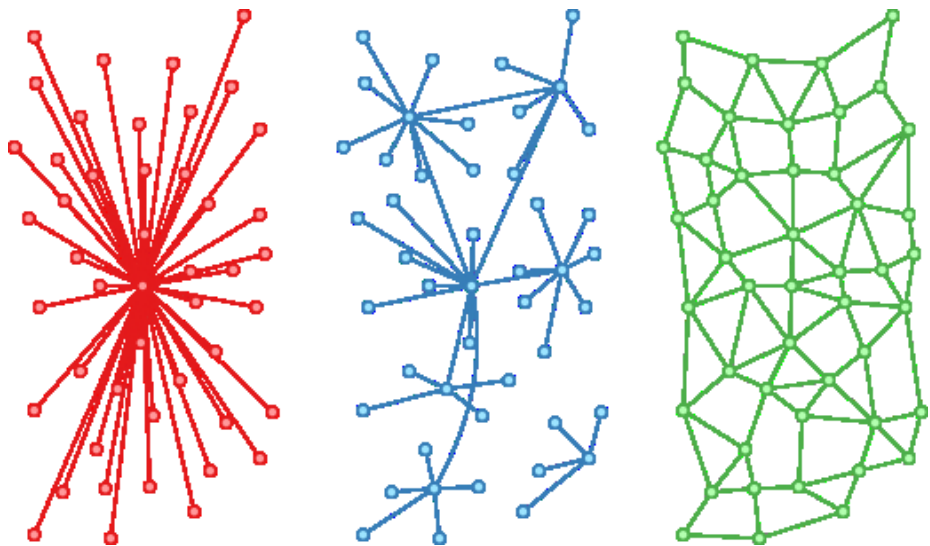}
\caption{From left to right: (1) Centralized (2) Federated (3) Decentralized}
\label{fig:baran}
\end{figure} 

\subsection{Centralized}

An architecture is strictly \emph{centralized} if there is a single trusted third-party that mediates all messages. Clients are restricted to a server. For example using a messaging service such as WhatsApp, for any two users, all messages must be passed through the WhatsApp server. Slack, Facebook, Signal, Apple Payments,  and other popular cloud-based services are examples of centralized architectures.

\subsection{Federated}

In the \emph{federated} architecture, client may pass messages to different servers, and these servers may communicate to pass messages between clients.  This is the classical architecture of the early Internet, where users were often offline and so servers were necessary. Open standards for protocols usually have this architecture, and so clients are not restricted to a single server, as otherwise server-mediated communication between multiple clients on different servers would be impossible. Taking e-mail for example, the user operates a client known as the Mail User Agent (MUA) that sends mail via one or more the Mail Transfer Agents (MTA, i.e. ``email servers'') until reaching the recipient's server. This server in turn sends the mail to the recipient's MUA. The security of the e-mail eco-system at least assumes each of these channels is encrypted via TLS ~\cite{foster2015security}.  SMTP services such as Gmail are federated, as would be any service using a standardized messaging protocol such as XMPP+OTR or even ``federated'' or ``permissioned'' blockchains. 

\subsection{Decentralized}

In the \emph{decentralized} model -- also a ``peer to peer'' (P2P) model) -- clients may directly communicate without a server, and so clients are considered ``peers''~\cite{peersoram}. In a peer-to-peer framework, clients may pass messages on behalf of other clients, although they may also drop messages. Servers are not used. This naturally leads to a default to broadcast messaging in peer-to-peer systems.  Earlier systems like Bittorrent as well as newer ``permission-less'' blockchain-based models fall into this category. 

 A proper subset of distributed systems that operate in an adversarial model are \emph{decentralized systems}, which are ``systems in which multiple authorities control different components and no single authority is fully trusted by all components''~\cite{troncoso2017systematizing}  As per malicious security~\cite{yung2015mobile}, both centralized and decentralized parties may no longer honestly follow the protocol, but may maliciously attempt to subvert the protocol: For example, a peer in a peer-to-peer system may drop traffic or claim falsely to execute part of the protocol while not doing so.

Distributed ledgers, and implementations such as blockchains, allow components of a system to record their activity transparently so that other peers can verify if components are following the protocol and malicious components detected~\cite{halpin:blockchain}. Distributed consensus protocols are then required to when the components need to have a consistent global view of some part of the system~\cite{lamport1982byzantine}, such as a distributed ledger of transactions. Distributed systems that do not require the quorum be set in advance but allow any peer to join are permissionless, and so decentralized. In contrast, if a quorum of existing members or other manual procedure is required to allow a new peer to enter the system, then the manual procedure is a centralized root of trust or the quorum of ``trusted'' peers equivalent to servers in the federated model in practice. 

\section{Analysis}
\label{sec:analysis}

The decentralization trilemma\footnote{\url{https://github.com/ethereum/wiki/wiki/Sharding-FAQ}} states that a system can only have two of three in terms of security, scalability, or decentralization but not all three. On one hand, a system may be both secure and scalable, but not decentralized. Google and other cloud-based services fit in this requirement. In contrast, a system may be decentralized and scalable, but not secure. Systems such as Bittorrent fall in this category~\cite{WolchokHHFHRWW10}. Finally, a system may be decentralized and secure, but not scalable. Blockchains such as Ethereum and Bittorrent would be part of this category.

Federated systems attempts to balance the trilemma by having multiple servers, but in reality this simply spreads the trust from one server to multiple servers. Therefore, federated systems are not ``trustless'' and so not decentralized, even if they use a blockchain. Some of the examples are subtle: Napster did use a peer-to-peer protocol, but had a centralized search directory. So Napster would count as federated, as the centralized directory server was required for the function of searching torrents, and each peer needs to connect to that server in order to function properly.

By tying the decentralization trilemma to concrete technical properties, we can determine if it actually holds. Every architecture makes certain design choices that privilege one property over another. Although more work is necessary in order to formally characterize each of these architectures, history suggests that these architectures are structurally biased toward certain properties and against others. Table \ref{tab:archcomp} provides a comparison of the choices made by the architectures outlined above.

\begin{table*}
\begin{center}
\begin{tabular}{|l|c|c|c|} \hline\hline
  \textbf{Property} & \emph{Centralized} & \emph{Federated} & \emph{Decentralized} \\ \hline\hline
\multicolumn{4}{|c|}{\textbf{Security and Privacy Properties}}        \\ \hline  
\emph{Confidentiality} & \cmark & \cmark & \cmark \\ \hline
\emph{Integrity} 	& \cmark & \cmark & \cmark \\ \hline
\emph{Authenticity} 	& \xmark & \xmark & \cmark \\ \hline
\emph{Unlinkability} 	& \xmark & \xmark & \cmark \\ \hline
\emph{Unobservability} 	& \xmark & \xmark & \xmark \\  \hline
\multicolumn{4}{|c|}{\textbf{Scalability Properties}}   \\ \hline 
\emph{Availability} 	& \cmark & \cmark & \xmark \\ \hline
\emph{Capacity}       & \cmark & \xmark & \xmark \\ \hline
\emph{Latency}        & \cmark & \cmark & \xmark  \\ \hline
\emph{Transparency}   & \xmark  & \xmark & \cmark*\\ \hline
\emph{Portability} 	& \xmark & \cmark & \xmark  \\ \hline \hline\hline
Examples: 	& Apple Pay, Signal & Visa, E-mail (PGP) & Bitcoin, Bittorrent \\
\hline\hline
\end{tabular}
\end{center}
\caption{Properties compared to architectures (* assumes a distributed ledger)}
\label{tab:archcomp}
\end{table*}

\subsection{Security Analysis}

\subsubsection{Confidentiality, Integrity, and Authenticity}
In general, all architectures are able to implement classical security properties using cryptography. However, authentication is a problem also for all architectures, albeit in different ways. Centralized architectures have low authenticity insofar as a user can only communicate to other users via a server and a server could serve false keys or impersonate its users, as is possible even with Signal~\cite{kobeissi2017automated}. This critique also applies to federated servers, although out-of-band finger-print verification may help, despite users often being unable to use fingerprint verification to detect a man-in-the-middle attack by a malicious server~\cite{schroder2016signal}. Peer to peer systems suffer from similar issues, as key discovery can be easily manipulated by a malicious peer. However, decentralization at least allows  direct communication of key material and non-custodial key ownership, and so decentralized systems have an advantage as a (possibly malicious) server is not required to mediate key material. Also, decentralization requiring each client to serve as a peer leads to an inability to cover certain use-cases, as security properties need to be enforced over groups, but each peer is meant to be an individual. Security operations over groups are more well-defined but much less commonly deployed~\cite{ChaumH91}. For example,  Bitcoin has issues with large group signatures and secure group messaging is still considered a hard problem, even in centralized environments~\cite{GoldbergUGC09}.

\subsubsection{Privacy: Unlinkability and Unobservability}
Without advanced techniques like differential privacy or secure multi-party computation, centralization has low unlinkability insofar as the malicious server can link any of its users to messages and can always map the network of a user via observing their messages. Likewise for authenticity issues in centralized systems, and the same is possible in federated settings. Decentralized systems by design can offer possibly more privacy as it is less trivial for a central server to monitor all transactions, although not without cost.  For example, even Tor is vulnerable to traffic analysis attacks on the exit and entrance nodes.~\cite{dingledine2004tsg}. Some decentralized designs allow a high amount of unlinkability, using techniques like zero-knowledge proofs such as ZCash~\cite{zcash} and unobservability via the use of cover traffic~\cite{ClarkeSWH00}, although these designs are not widely deployed. The problem is cover traffic and other techniques to increase anonymity come at the cost of reducing the capacity of the network, as given by the well-known ``anonymity trilemma'' between capacity, latency, and anonymity~\cite{das2018anonymity}.

\subsection{Scalability Analysis}

\subsubsection{Availability, Capacity, and Latency}

Scalability concerns the reliable operations of the system in the face of growing demand. Due to their use of robust distributed (but not decentralized) systems, centralized architectures are highly available and can easily scale to having a high capacity with low latency~\cite{dynamo}. Federated systems have as a bottle-neck the server with the least capacity (similar to decentralized systems). In most decentralized architectures, availability is low as peers may always be offline. Although federation may be a disadvantage in terms of unlinkability in comparison to decentralized systems, federated systems typically have much higher availability than decentralized designs as the server is normally online while peer-to-peer systems suffer from network churn.  This tends to translate into lower latency in comparison to decentralized systems. In centralized systems, it is much easier to upgrade capacity via simply upgrading servers and connection capacity. In decentralized systems, the problem of the ``weakest'' link of capacity can be profound, particularly if there is no way to route around low capacity peers. Therefore, real-world decentralized systems like Bittorrent tend to evolve ``super-nodes'' that have high capacity and are online. These ``super nodes'' in effect emerge as de-facto ``servers'' in the decentralized system~\cite{WangK13}, undermining the security and privacy properties of decentralized systems and making them de-facto federated systems. 

\subsubsection{Transparency and Portability}

Centralized and federated servers whose code and operations are hidden from their users on a server are difficult for third parties to audit by their users, and so transparency is low. Decentralized systems offer less ability for an adversary to tamper with incoming messages as the malicious server has been eliminated insofar as messages may be directly sent from user to user, although in practice a relaying peer may tamper with the messages. For example,  Bittorrent has no transparent log, and so nodes dropping packets cannot be easily detected.

 Although decentralization gets rid of a malicious server, clients may join and act maliciously in a \emph{sybil} attack~\cite{Douceur02}. Therefore, blockchain technology requires enforcing transparency in a decentralized setting so attacks and failures by peers can be discovered. In terms of portability, decentralized systems suffer from a dearth of interoperable standards, while federated systems are highly standardized (e-mail, XMPP, etc.). The usability of most peer-to-peer systems tend to be poor, while federated systems such as email have been able to compete in terms of usability with centralized systems.

 \section{Open Problems}
 \label{sec:problems}
 
Our analysis leads to a number of open problems exist that prevent the further development of decentralized alternatives.  Some of the disadvantages of decentralization can be solved by standardization in order to bring the portability of decentralized architectures in competition with well-known federated protocols. Better usability is desperately needed to drive more users from centralized to decentralized systems. Some of the characteristics of architectures could be changed by technical advances: For example, increased usage of cover traffic could help address unobservability when combined with mix networking~\cite{DanezisDTL10}. Yet some problems are more fundamental. In surveys of secure communication \cite{unger2015sok}, a pattern starts to emerge: Any serious attempt to build a decentralized alternative that is resistant to a malicious server eventually comes up against similar hard problems.

It is possible to ignore many of these problems if one rules out one of these properties, but users  have grown accustomed to contemporary methods of online communication based on silos  (Facebook, Twitter, etc.) that are easy to use and have high availability, as well as assume a level of authenticity that is normally not justified by the difficulties of trusted key discovery in centralized systems. So, decentralized architectures that are designed to defend against malicious servers should find solutions to the ``seven hard problems'' as follows:\footnote{An earlier version of this list was originally defined at LEAP: \url{https://leap.se/en/docs/tech/hard-problems}} 

\begin{enumerate}
\item \textbf{Public key discovery problem}: Public key discovery and validation is very difficult for users to manage in a decentralized system, but without it a message cannot have authenticity.
\item \textbf{Key Availability problem}: Users would like communicate seamlessly and securely across different devices, as well as restore data if a device or key is lost.
\item \textbf{Group problem}: Users work in social groups, yet public key cryptography does not have a consistent manner for dealing with groups in decentralized settings. This effects confidentiality, integrity, and authenticity.
  \item \textbf{Metadata problem}: Messages are vulnerable to traffic analysis, unless cover traffic is used, which is not common in decentralized systems. This derives from issues around unlinkability and unobservability.
   \item \textbf{Sybil problem}: Without servers, any client may join a decentralized network and act maliciously in a sybil attack~\cite{Douceur02}.
   \item \textbf{Update problem}: Software updates need to be securely delivered across multiple clients in a decentralized system, which is easy to do with a centralized server.
   \item \textbf{Resource problem}: It is an open problem on how to let users securely share a resource for real-time collaboration in a decentralized architecture.
\end{enumerate}

These problems cause real-world impact on interest in decentralized architectures both by users and researchers. New standards like IETF Message Layer Security seem to be allowing encrypted user communication to tackle the \emph{group problem}, but it is not clear how to apply to decentralized systems.\footnote{\url{https://datatracker.ietf.org/wg/mls/}} The \emph{update problem} is partially addressed in terms of security by reproducible builds. The \emph{resource problem} in part is due to the group problem being unsolved in tandem with scalability issues, where decentralized designs have a lack of availability and latency that make sharing a single resource between a group difficult.

%Open issues in terms of software updates and group-access to resources though, mean that currently cloud-based systems still better address the needs of users, even if there are potential advantages in terms of security for blockchain systems in the face of malicious servers.

\section{Conclusions}
\label{sec:end}

The decentralization trilemma is a simplification that can be deconstructed into traditional security and scalability properties. Nonetheless, decentralization solves a real problem: Decentralization prevents a single centralized malicious server from compromising the security of users. There is a kernel of truth in the decentralization trilemma: In broad strokes, decentralization does offer better security at the cost of scalability if end-users manage their keys and can offer higher scalability if sacrificing security is acceptable. Nonetheless, decentralized systems naturally have scaling issues as nodes are more frequently off-line or faulty. Worse, communication can require more hops than in centralized systems, increasing the chance of failure. 

The real breakthrough in blockchain technology, in comparison to earlier peer-to-peer decentralized systems, is the use of distributed ledgers to prevent malicious behavior by peers via transparency. This naturally leads to loss of latency due to the nature of distributed consensus protocols~\cite{lamport1978time}.  Blockchain technology, while not solving the fundamental issues around availability that plague decentralized systems, does allow malicious servers to be removed and malicious peers to be detected, and could lessen the security problems caused by ``super nodes.'' If peers could be ubiquitously online, they should be more available, which may increase capacity  and decrease latency. Yet to do so securely would mean the network would reach widespread usage and users would be highly security-conscious, which appears delusional as most users are not systems administrators.

  The low availability of decentralized architectures likely necessitates learning lessons about the evolution of ``super-nodes'' (high capacity nodes that relay the majority of the traffic) in peer-to-peer architectures, which can act as high availability relays and so bring the benefits of federated designs to decentralized systems. In order to secure these ``super-nodes,'' it does seem to make sense that a specialized class of system administrators would arise in federated systems. 

Thus, the technical problem of decentralization is revealed at its core  to be a political problem: Centralization to a large extent arises due to the monopoly of technical knowledge by a class of system administrators and programmers. The goal of decentralized systems seems to be to spread this technical knowledge so that all users can autonomously operate and govern their own infrastructure. In this way, federated and centralized models represent the inherent trust assumptions of non-technical users. Historically, humans typically trust friends and associates, as well as deploy a specialization of labor. So a user would likely trust a highly-skilled individual they know in some fashion, perhaps a friend or affiliate at a human-scale institution like a university, to manage technical infrastructure on their behalf. The rise of a non-profit federated systems at the dawn of the Internet maps to pre-existing human trust relationships. As the technical complexity of the Internet scaled and users needed more convenience, this model shifted as users trusted Google to handle the sheer technical complexity in return for profit off of personal data or for a certain cost. The lack of trust in these centralized servers and the humans behind them is what led to the rise of peer-to-peer and blockchain: Decentralization is a political ideology masquerading as a technical architecture.

The next steps to explore this thesis would require formalizing properties such as availability, latency, and capacity in a more rigorous manner than presented here, similar as what has been done in research on privacy~\cite{das2018anonymity}. Then architectures could be compared with the same rigor as security properties within a formal message-passing framework as done by the $\pi$-calculus~\cite{kobeissi2017automated}. Another promising avenue is using network-theoretic approaches to study the evolution of decentralized networks into super-nodes. Further socio-technical work in needed in understanding the motivations, usability, and political stakes of decentralization. The ideology of decentralization has given us very real advances in technology for ordinary users, but much more is needed for liberation.

%\textbf{Acknowledgements}: The authors are supported by NEXT\-LEAP (EU H2020 ref: 688722) and the Open Technology Fund.

\bibliographystyle{apalike}
{\small \bibliography{main}}

\begin{thebibliography}{}

\bibitem[Baran et~al., 1964]{baran1964distributed}
Baran, P. et~al. (1964).
\newblock On distributed communications.
\newblock {\em Volumes I-XI, RAND Corporation Research Documents, August}.

\bibitem[Ben{-}Sasson et~al., 2014]{zcash}
Ben{-}Sasson, E., Chiesa, A., Garman, C., Green, M., Miers, I., Tromer, E., and
  Virza, M. (2014).
\newblock Zerocash: Decentralized anonymous payments from bitcoin.
\newblock In {\em {IEEE} Symposium on Security and Privacy}.

\bibitem[Chaum and van Heyst, 1991]{ChaumH91}
Chaum, D. and van Heyst, E. (1991).
\newblock Group signatures.
\newblock In {\em Advances in Cryptology}.

\bibitem[Clarke et~al., 2000]{ClarkeSWH00}
Clarke, I., Sandberg, O., Wiley, B., and Hong, T.~W. (2000).
\newblock Freenet: {A} distributed anonymous information storage and retrieval
  system.
\newblock In {\em Designing Privacy Enhancing Technologies, International
  Workshop on Design Issues in Anonymity and Unobservability}.

\bibitem[Danezis et~al., 2010]{DanezisDTL10}
Danezis, G., D{\'{\i}}az, C., Troncoso, C., and Laurie, B. (2010).
\newblock Drac: An architecture for anonymous low-volume communications.
\newblock In {\em 10th Privacy Enhancing Technologies Symposium}.

\bibitem[Das et~al., 2018]{das2018anonymity}
Das, D., Meiser, S., Mohammadi, E., and Kate, A. (2018).
\newblock Anonymity trilemma: Strong anonymity, low bandwidth overhead, low
  latency-choose two.
\newblock In {\em 2018 IEEE Symposium on Security and Privacy (SP)}, pages
  108--126. IEEE.

\bibitem[DeCandia et~al., 2007]{dynamo}
DeCandia, G., Hastorun, D., Jampani, M., Kakulapati, G., Lakshman, A., Pilchin,
  A., Sivasubramanian, S., Vosshall, P., and Vogels, W. (2007).
\newblock Dynamo: Amazon's highly available key-value store.
\newblock In {\em ACM Symposium on Operating Systems Principles: Proceedings of
  twenty-first ACM SIGOPS symposium on Operating systems principles},
  volume~14, pages 205--220.

\bibitem[Dingledine et~al., 2004]{dingledine2004tsg}
Dingledine, R., Mathewson, N., and Syverson, P. (2004).
\newblock {Tor: The second-generation onion router}.
\newblock {\em Proceedings of the 13th USENIX Security Symposium}, 2.

\bibitem[Douceur, 2002]{Douceur02}
Douceur, J.~R. (2002).
\newblock The sybil attack.
\newblock In {\em 1st International Worksop on Peer-to-Peer Systems}.

\bibitem[Foster et~al., 2015]{foster2015security}
Foster, I.~D., Larson, J., Masich, M., Snoeren, A.~C., Savage, S., and
  Levchenko, K. (2015).
\newblock Security by any other name: On the effectiveness of provider based
  email security.
\newblock In {\em Proceedings of the 22nd ACM SIGSAC Conference on Computer and
  Communications Security}, pages 450--464. ACM.

\bibitem[Goldberg et~al., 2009]{GoldbergUGC09}
Goldberg, I., Ustaoglu, B., Gundy, M.~V., and Chen, H. (2009).
\newblock Multi-party off-the-record messaging.
\newblock In {\em 16th {ACM} Conference on Computer and Communications
  Security}.

\bibitem[Greenwald, 2014]{greenwald:noplace}
Greenwald, G. (2014).
\newblock {\em {No Place to Hide: Computer Hacking, Crashing, Pirating, and
  Phreaking}}.
\newblock Metropolitan Books.

\bibitem[Halpin and Piekarska, 2017]{halpin:blockchain}
Halpin, H. and Piekarska, M. (2017).
\newblock Introduction to security and privacy on the blockchain.
\newblock In {\em 2017 IEEE European Symposium on Security and Privacy
  Workshops (EuroS\&PW)}, pages 1--3. IEEE.

\bibitem[Kobeissi et~al., 2017]{kobeissi2017automated}
Kobeissi, N., Bhargavan, K., and Blanchet, B. (2017).
\newblock Automated verification for secure messaging protocols and their
  implementations: A symbolic and computational approach.
\newblock In {\em 2017 IEEE European Symposium on Security and Privacy
  (EuroS\&P)}, pages 435--450. IEEE.

\bibitem[Lamport, 1978]{lamport1978time}
Lamport, L. (1978).
\newblock Time, clocks, and the ordering of events in a distributed system.
\newblock {\em Communications of the ACM}, 21(7).

\bibitem[Lamport et~al., 1982]{lamport1982byzantine}
Lamport, L., Shostak, R., and Pease, M. (1982).
\newblock The byzantine generals problem.
\newblock {\em ACM Transactions on Programming Languages and Systems (TOPLAS)},
  4(3):382--401.

\bibitem[Minar and Hedlund, 2001]{peersoram}
Minar, N. and Hedlund, P. (2001).
\newblock A network of peers – peer-to-peer models through the history of the
  internet.
\newblock In Oram, A., editor, {\em Peer-to-Peer: Harnessing the Power of
  Disruptive Technologies}, pages 9--20. O’Reilly, Sebastopol, CA.

\bibitem[Pfitzmann and Hansen, 2005]{PfitzmannH05}
Pfitzmann, A. and Hansen, M. (2005).
\newblock Anonymity, unlinkability, unobservability, pseudonymity, and identity
  management -- a consolidated proposal for terminology.
\newblock Technical report.

\bibitem[Schr{\"o}der et~al., 2016]{schroder2016signal}
Schr{\"o}der, S., Huber, M., Wind, D., and Rottermanner, C. (2016).
\newblock When signal hits the fan: on the usability and security of
  state-of-the-art secure mobile messaging.
\newblock In {\em European Workshop on Usable Security. IEEE}.

\bibitem[Slobogin, 2014]{slobogin2014cause}
Slobogin, C. (2014).
\newblock Cause to believe what?: The importance of defining a search's
  object--or, how the aba would analyze the nsa metadata surveillance program.
\newblock {\em Oklahoma Law Review}, pages 14--7.

\bibitem[Troncoso et~al., 2017]{troncoso2017systematizing}
Troncoso, C., Isaakidis, M., Danezis, G., and Halpin, H. (2017).
\newblock Systematizing decentralization and privacy: Lessons from 15 years of
  research and deployments.
\newblock {\em Proceedings on Privacy Enhancing Technologies},
  2017(4):404--426.

\bibitem[Unger et~al., 2015]{unger2015sok}
Unger, N., Dechand, S., Bonneau, J., Fahl, S., Perl, H., Goldberg, I., and
  Smith, M. (2015).
\newblock Sok: Secure messaging.
\newblock In {\em 2015 IEEE Symposium on Security and Privacy}, pages 232--249.
  IEEE.

\bibitem[Wang and Kangasharju, 2013]{WangK13}
Wang, L. and Kangasharju, J. (2013).
\newblock Measuring large-scale distributed systems: case of {BitTorrent}
  mainline {DHT}.
\newblock In {\em 13th {IEEE} International Conference on Peer-to-Peer
  Computing}.

\bibitem[Wolchok et~al., 2010]{WolchokHHFHRWW10}
Wolchok, S., Hofmann, O.~S., Heninger, N., Felten, E.~W., Halderman, J.~A.,
  Rossbach, C.~J., Waters, B., and Witchel, E. (2010).
\newblock Defeating vanish with low-cost sybil attacks against large dhts.
\newblock In {\em Network and Distributed System Security Symposium}.

\bibitem[Yung, 2015]{yung2015mobile}
Yung, M. (2015).
\newblock The mobile adversary paradigm in distributed computation and systems.
\newblock In {\em Proceedings of the 2015 ACM Symposium on Principles of
  Distributed Computing}, pages 171--172. ACM.

\end{thebibliography}
\end{document}